# STOCHASTIC COHERENCE IN AN OSCILLATORY GENE CIRCUIT MODEL

Robert C. Hilborn[a,*] and Jessie D. Erwin[b]

[a]*Department of Physics, The University of Texas at Dallas, Richardson, TX 75082*

[a,b]*Department of Physics, Amherst College, Amherst, MA 01002*

**Abstract**

We show that noise-induced oscillations in a gene circuit model display stochastic coherence, that is, a maximum in the regularity of the oscillations as a function of noise amplitude. The effect is manifest as a system-size effect in a purely stochastic molecular reaction description of the circuit dynamics. We compare the molecular reaction model behavior with that predicted by a rate equation version of the same system. In addition, we show that commonly used reduced models that ignore fast operator reactions do not capture the full stochastic behavior of the gene circuit. Stochastic coherence occurs under conditions that may be physiologically relevant.

[*] Corresponding author. Science/Mathematics Education Department, The University of Texas at Dallas, 800 W. Campbell Road, FN33, Richardson, TX 75080. Tel: +1 972 883 4726; fax +1 972 883 6976. *Email address*: rhilborn@utdallas.edu

## 1. Introduction

Gene circuits consist of DNA operator and coding regions that control the production of particular proteins and the proteins themselves (and perhaps other regulatory molecules), which may act as activators or repressors for DNA transcription, both at the original site and at other sites. We shall refer to the combination of operator site(s) and associated coding region(s) as a gene. The gene copy number is the number of genes of a particular type in a cell. Many time-dependent biological functions, including circadian rhythms, are regulated by the dynamics of gene circuits (Barkai and Leibler, 2000). The periodicity of this behavior is somewhat surprising because the gene copy number and the number of resulting proteins are often quite small in a given cell, so one would expect substantial relative fluctuations in the number of proteins at any particular DNA operator site and hence considerable fluctuations in the reaction rates (Guido et al., 2006; Kærn et al., 2005; McAdams and Arkin, 1997; Mettetal et al., 2006). More generally, stochastic effects in gene expression may be responsible for cell-to-cell variations seen in populations with identical genomes (Elowitz et al., 2002). Other studies have examined the propagation of noise in gene circuits (Pedraza and van Oudenaarden, 2005).

In this paper we focus on the effects of fluctuations on the dynamics of an oscillatory gene circuit in which the number of proteins produced is strongly time-dependent. We demonstrate that fluctuation-driven oscillations in the model gene circuit can display a high degree of regularity and, in fact, the regularity has a maximum as a function of noise amplitude, an effect known as stochastic coherence (Zaikin et al., 2003) (also called coherence resonance (Pikovsky and Kurths, 1997)). In particular, we show that a purely stochastic description of the gene circuit dynamics exhibits a regularity maximum as a function of system size (the copy number). Previous models of stochastic coherence (in systems other than gene circuits) have used differential equation (rate equation) formulations, whose validity is questionable in the context of gene networks given that only a small number of active molecules are present for a typical gene circuit. The regularity of the oscillations in a complete oscillatory gene circuit (including the operator sites, coding regions, messenger RNA, transcription, translation, and protein degradation) has not been studied before as a function of the amplitude of the fluctuations. Other authors (Hou and Xin, 2003; Wang et al., 2007; Wang et al., 2005) have studied the effects of noise on simplified models of oscillatory gene circuits, but those models did not include the operator dynamics and the "optimal" noise ranges did not correspond to physiologically relevant regimes. We show that such reduced models (models that ignore the fast operator dynamics) miss important aspects of the stochastic behavior.

## 2. The VKBL Model

Most gene networks (combinations of many gene circuits) in nature are quite complex. See, for example, (Becker-Weimann et al., 2004). Therefore, we focus our attention on a simplified model (Barkai and Leibler, 2000; Vilar et al., 2002) of an oscillatory gene circuit that captures the basic features of more complex systems. The model consists of two distinct DNA sequences that lead to the production of two proteins, Activator and Repressor. The Activator protein A can bind to the so-called operator regions on the two DNA sequences, and its presence at those operator sites significantly enhances the rate of transcription of the DNA into messenger-RNA (mRNA), which is then translated to form the proteins. The Repressor protein R can bind with Activator to form a protein complex that effectively keeps Activator from binding to the operator sites. The resulting gene circuit can be described by the following sequence of 16 reactions:

$$D_A + A \xrightarrow{\gamma_A} D'_A \tag{1}$$

$$D'_A \xrightarrow{\theta_A} D_A + A, \tag{2}$$

$$D_R + A \xrightarrow{\gamma_R} D'_R \quad (3)$$

$$D'_R \xrightarrow{\theta_R} D_R + A \quad (4)$$

$$D'_A \xrightarrow{\alpha'_A} D'_A + M_A \quad (5)$$

$$D_A \xrightarrow{\alpha_A} D_A + M_A \quad (6)$$

$$M_A \xrightarrow{\delta_{M_A}} X \quad (7)$$

$$M_A \xrightarrow{\beta_A} M_A + A \quad (8)$$

$$A + R \xrightarrow{\gamma_C} C \quad (9)$$

$$A \xrightarrow{\delta_A} Y \quad (10)$$

$$D'_R \xrightarrow{\alpha'_R} M_R + D'_R \quad (11)$$

$$D_R \xrightarrow{\alpha_R} M_R + D_R \quad (12)$$

$$M_R \xrightarrow{\delta_{M_R}} Z \quad (13)$$

$$M_R \xrightarrow{\beta_R} M_R + R \quad (14)$$

$$C \xrightarrow{\delta_A} R \quad (15)$$

$$R \xrightarrow{\delta_R} W \quad (16)$$

In Eqs. (1)-(16), the non-italic symbols represent the specific molecule type (rather than the number of molecules). $D'_A$ and $D_A$ represent the DNA operator sites with and without protein A (Activator) bound, respectively. $D'_R$ and $D_R$ are the corresponding R (Repressor) protein DNA operator sites. $M_A$ and $M_R$ are the mRNAs for the two proteins. C represents the Activator-Repressor complex. The model assumes that when the complex decomposes Activator is degraded. Thus, $\delta_A$ appears in Eq. (15). W, X, Y, and Z are inactive decay products. The model ignores any changes in concentration due to cell growth or cell division (mitosis).

We use the following values for the rate constants (Vilar et al., 2002): $\theta_A = 50$, $\theta_R = 100$, $\alpha_A = 50$, $\alpha'_A = 500$, $\alpha_R = 0.01$, $\alpha'_R = 50$, $\delta_{M_A} = 10$, $\delta_{M_R} = 0.5$, $\delta_A = 1$, $\beta_A = 50$, $\beta_R = 5$, and $\delta_R = 0.06$, all in hour$^{-1}$. The parameters $\gamma_A = 1$, $\gamma_R = 1$, and $\gamma_C = 2$ are given in molecules$^{-1}$ hour$^{-1}$. With these rate constants, the transcription efficiency (that is, the average number of mRNAs produced between successive operator activations) (Kærn et al., 2003) is 10 for Activator-mRNA and 0.5 for Repressor-mRNA. The translation efficiency (the average number of proteins per mRNA, sometimes called the "burst size") is 5 for Activator and 10 for Repressor. The model was designed to capture the general features of genetic oscillators, and the parameters, though rarely known in detail for natural gene circuits, are typical values for circadian oscillators (Becker-Weimann et al., 2004; Vilar et al., 2002). Following Vilar et al. (2002), we use $\delta_R$, the degradation rate of Repressor, as a control parameter. (Recent work indicates that the degradation of cryptochrome proteins controls circadian clock oscillations in mammals (Busino et al., 2007).) We now turn to a description of the dynamics given by the reactions. It is crucial to note that in most cells, there exist only one or a few copies of the relevant genes (Raser and O'Shea, 2005). Therefore any study of the dynamics ought to be based on discrete, stochastic dynamics. The dynamics of the model can be simulated using a stochastic reaction Monte Carlo algorithm developed by Gillespie (Gillespie, 1976; Gillespie, 1977). In the language of stochastic processes, the Gillespie algorithm treats the various biochemical processes as "one-step" or "birth-death" processes (van Kampen, 1992). Gillespie (1977) argued that under a wide range of conditions this algorithm provides an "exact" model of the reaction dynamics.

The Gillespie algorithm proceeds as follows: the time $\tau$ between subsequent reactions is determined by drawing a random number $0 < r_1 < 1$ and then calculating $\tau = (1/a_0)\ln(1/r_1)$, where $a_0$ is the sum of all the reaction rates, which include the appropriate combinatorial factors for the number dependence of the reacting species. This scheme provides an exponentially-distributed waiting time between reactions. The particular reaction to be executed is found by selecting a second random number $0 < r_2 < 1$ and finding the largest reaction number $j$ that satisfies $\sum_{i=1}^{j-1} a_i < r_2 \times a_0$. The species numbers are then updated for that particular reaction. This two-step cycle is continued until all of the molecules disappear, a specified number of reactions has occurred, or a specified time has passed. Figure 1 illustrates the fluctuation-induced oscillations of the number of Repressor proteins for the stochastic reaction model with gene copy numbers $N_A = N_B = 1$ for conditions for which the deterministic model (to be discussed below) yields only fixed point behavior.

We look for stochastic coherence in the stochastic reaction model of the dynamics by varying the copy numbers $N_A$ and $N_R$ for both the Activator and Repressor proteins, which will control the average number of Repressor and Complex proteins between the pulses. (Between pulses the number of free Activators drops close to 1.)

To characterize the degree of periodicity of the dynamics, we use the regularity, R, defined as

$$\mathsf{R} = \frac{\langle T \rangle}{\sqrt{\mathrm{var}(T)}}, \qquad (17)$$

where $T$ is the time between subsequent protein pulses and var($T$) is the variance of the time intervals. The angle brackets indicate a time average. R is just the reciprocal of the coefficient of variation commonly used to characterize the statistics of interspike intervals in neurons (Dayan and Abbott, 2001).

Figure 2 displays the regularity of the fluctuation-induced pulses as a function of the logarithm of the gene copy number, where we have used $N_A = N_R$. The results exhibit stochastic coherence; that is, there is a maximum in the regularity as a function of the gene copy number. The maximum in the regularity observed here is analogous to "system size" stochastic coherence (coherence resonance) observed in models of calcium release in cells (Jung and Shuai, 2001; Schmid et al., 2001; Shuai and Jung, 2002a; Shuai and Jung, 2002b; Zhang et al., 2004) and neural action potentials (Shuai and Jung, 2005; Zeng and Jung, 2004). The important difference is that here the fluctuations are due entirely to fluctuations in the reactions and the number of molecules. Previous studies (Jung and Shuai, 2001; Schmid et al., 2001; Shuai and Jung, 2002a; Shuai and Jung, 2002b; Shuai and Jung, 2005; Zeng and Jung, 2004) of system-size effects in calcium signaling and in neurons have used hybrid dynamics with stochastic differential equations for concentrations or membrane voltages and discrete stochastic processes for channel openings and closings.

Several papers (Hou and Xin, 2003; Wang et al., 2007; Wang et al., 2005) have reported investigations of the effects of "internal" noise (due to fluctuations in the relatively small numbers of molecules) in gene clock systems using the stochastic reaction description. However, the models used in those papers ignored the operator activation and de-activation dynamics and consequently, the interesting noise behavior occurred in regimes that do not seem to be physiologically relevant. The assumption employed implicitly in those papers was that the operator activation dynamics and de-activation dynamics are fast compared to the transcription, translation, and degradation dynamics and hence could be "adiabatically" eliminated. Our results, to be described below, indicate that this assumption, which may be valid for systems with large numbers of constituents, misses important contributions to the stochastic behavior of gene circuits with small copy numbers.

Figure 2 also shows the effects of changing the transcription rates indicated by the αs in Eqs. (1)-(16). With larger transcription rates, the numbers of Repressor and Activator proteins produced are increased. As the data in Fig. 2 indicate, there are no dramatic changes in the regularity over this parameter range. The location of the maximum regularity seems to shift to larger values of gene copy number as the transcription rate decreases, but that shift is just at the edge of statistical significance. It is interesting to note that the maximum occurs in the range of two-four gene copies.

If the gene copy number is small and held fixed, increasing the transcription rates leads to an increase in the number of Activator and Repressor proteins. However, the system behavior does not approach that of the deterministic rate equations (a steady state, as described below) because the relative fluctuations in the number of active DNA operator sites remains large, and those fluctuations continue to induce oscillations. In fact, the regularity increases with Repressor protein numbers but then saturates (at about 7.8 for the conditions given in Fig. 2). In other words, the effects due to the small number of gene copies dominate the stochastic dynamics. Why is there a maximum in the regularity as a function of the gene copy number? The explanation is as follows: In the absence of fluctuations, for the parameter values used here, the dynamics of the model tend to a steady-state after transients die away. (See the rate equation analysis below.) The system must wait near the steady state conditions until a sufficient large noise fluctuation pushes the system far enough to induce a "pulse" (or burst) of proteins. When the gene copy number is small, the time between bursts is more irregular and the resulting value of the regularity is small. As the gene copy number increases, the noise level increases (although the relative noise decreases) and the bursts occur as soon as the system returns to the neighborhood of the steady-state fixed point. Since the duration of the burst is determined primarily by the decay of Repressor, the regularity of the bursts increases. Finally, when the gene copy number becomes sufficiently large, the relative fluctuations diminish and large (relative) fluctuations become rarer; the rate of bursting decreases and the interval between bursts becomes more irregular, thereby decreasing the regularity. The combination of these effects leads to a maximum in the regularity as a function of gene copy number.

## 3. Rate Equations

The VKBL model can also be described by nine rate equations for the number of bound and unbound operator sites (indicated by italicized symbols) and accompanying coding regions, mRNAs, and the resulting proteins and the protein complex. Following the notation of Vilar et al. (2002), we write the rate equations as

$$dD_A / dt = \theta_A D'_A - \gamma_A D_A A \tag{18}$$

$$dD_R / dt = \theta_R D'_R - \gamma_R D_R A \tag{19}$$

$$dD'_A / dt = \gamma_A D_A A - \theta_A D'_A \tag{20}$$

$$dD'_R / dt = \gamma_R D_R A - \theta_R D'_R \tag{21}$$

$$dM_A / dt = \alpha'_A D'_A + \alpha_A D_A - \delta_{M_A} M_A \tag{22}$$

$$dM_R / dt = \alpha'_R D'_R + \alpha_R D_R - \delta_{M_R} M_R \tag{23}$$

$$dA / dt = \beta_A M_A + \theta_A D'_A + \theta_R D'_R \\ - A(\gamma_A D_A + \gamma_R D_R + \gamma_C R + \delta_A) \tag{24}$$

$$dR / dt = \beta_R M_R - \gamma_C AR + \delta_A C - \delta_R R \tag{25}$$

$$dC / dt = \gamma_C AR - \delta_A C, \tag{26}$$

where the italic symbols indicate the number of molecules present of each type. The rate equation model assumes continuous variables with no stochastic effects.

The behavior of the oscillations predicted by the rate equations is significantly different from the results found for the stochastic reaction dynamics. As before, we use $\delta_R$ as the control parameter. For $\delta_R < 0.0878$, the solutions of Eqs. (18)-(26) settle to a stable fixed point. In particular, for $\delta_R = 0.06$, the rate equation behavior is a steady-state (after transients die out). As the gene copy number increases, the stochastic reaction behavior eventually approaches that described by the deterministic rate equations. For smaller gene copy numbers, however, the reaction fluctuations induce oscillations, as noted previously, even under conditions for which the deterministic rate equations predict a stable steady state. This result points out one of the limitations of the rate equation description of gene circuit behavior.

There are several ways to account for the effects of the fluctuations on the gene circuit dynamics in the rate equation version of the model (Kepler and Elston, 2001; Steuer, 2004). In the simplest approach (Gillespie, 2000; van Kampen, 1992), we add a stochastic term to one (or more) of Eqs. (18)-(26) to set up a Langevin-type equation. For example, a stochastic version of Eq. (25) can be written as

$$dR/dt = \beta_R M_R - \gamma_C AR + \delta_A C - \delta_R R + \sqrt{d_R R}\,\eta(t),\qquad(27)$$

where $d_R R$ is the variance of the noise and $\eta(t)$ is a Gaussian-distributed random process with zero mean and standard deviation equal to 1. We have made the noise term dependent on the number of molecules to mimic Poisson-distributed molecule number fluctuations (Steuer, 2004; van Kampen, 1992), though we note that such an ansatz is problematic (van Kampen, 1992). Various intracellular processes such as localization through binding and active transport may lead to non-Poisson statistics and $d_R$ allows us to adjust the noise dependence. In this paper we focus on the so-called intrinsic (internal) fluctuations (Elowitz et al., 2002; Mettetal et al., 2006) associated with the number fluctuations of the various molecular species. We ignore global ("extrinsic") fluctuations (Elowitz et al., 2002; Mettetal et al., 2006; Volfson et al., 2006) such as cell growth and cell-wide fluctuations in polymerases that may affect production rates and decay rates.

The qualitative features of the noise-induced oscillations are largely independent of the details of the fluctuation sources (Lindner et al., 2004). The quantitative details of the regularity of the noise-induced oscillations, however, do depend on how the stochastic terms are added to the rate equations. The greatest difference occurs in the results with noise added to the rate equation with the fastest dynamics compared to those with noise added to the slowest dynamics (Hilborn and Erwin, 2004; Hilborn and Erwin, 2005). Here we focus on the dynamics when noise is added to Eq. (24) or Eq. (25). In a subsequent publication, we shall discuss more general situations. This model falls into the class of excitable dynamical systems (Lindner et al., 2004). We set $\delta_R = 0.06$ so that the deterministic versions of Eqs. (18)−(26) have solutions that settle to a stable fixed point. Then, if a sufficiently large noise "kick" bumps the system away from the fixed point, the trajectory will undergo a large excursion (a pulse) through state space before returning to the fixed point. Figure 3 illustrates the dynamics of the system when we add a noise term to Eq. (24): the gene circuit exhibits noise-induced oscillations.

Figure 4 displays the regularity of the gene circuit behavior as a function of $d_A$, when noise with variance $d_A A$ is added only to Eq. (24) and as a function of $d_R$, when noise with variance $d_R R$ is added only to Eq. (25). The results display stochastic coherence: the regularity exhibits a maximum as a function of the noise variance. The existence of the maximum can be explained qualitatively, in analogy with the stochastic reaction model, as follows: The fluctuating term will occasionally kick the system far enough from the fixed point to allow for a large excursion (a pulse) through state space. For low noise values, the pulses occur randomly and the number of pulses per unit time interval is described by a Poisson distribution for which $\text{var}(T) = T^2$ and hence $R \approx 1$. As the noise variance increases, the average pulse rate increases, but the deterministic behavior of the pulse itself is little affected by the noise. These effects tend to increase the regularity of the sequence of pulses. Eventually, however, the noise variance is sufficiently large that even the pulse behavior becomes irregular and the regularity decreases. As a result of these two trends, the regularity exhibits a maximum as a function of noise variance. For the set of parameters used, the dynamics associated with *A* is fast compared to that of *R*. The details of stochastic coherence depend on the time-scale separation between fast and slow dynamics and whether the noise is added to the fast dynamics or the slow dynamics (Hilborn and Erwin, 2004; Hilborn and Erwin, 2005). The results shown in Fig. 4 provide an example of the fast/slow dynamics difference. This distinction between fluctuations in fast and slow variables has been demonstrated previously in stochastic differential models of neural dynamics (Hilborn and Erwin, 2004; Hilborn and Erwin, 2005). Hilborn and Erwin (2005) give a detailed theoretical treatment of this effect in the context of a model of an excitable neuron.

**4. An adiabatic model**

Many models of genetic networks assume that the activation and de-activation of the promoter sites or enzyme dimerization are very rapid compared to other genetic events such as transcription, translation, and protein degradation. If that assumption holds, the number of activator and de-activated sites will be in a quasi-steady state depending on the concentration of the activator proteins. Alternatively, for low copy number situations, we can think of a steady-state fraction of the time during which the promoter is activated. That fraction depends on the concentration of the activator protein. In either case, we can then eliminate those activation and de-activation events from the analysis of the genetic network dynamics. In standard physics terminology, we are performing adiabatic elimination of the fast (rapidly varying) variables. We now explore the effects of adiabatic elimination on stochastic coherence in the VKBL model. First, we apply these notions to the rate equation version of the VKBL model. We focus on the activation and de-activation events and the resulting transcription rates described by Eqs.(18)-(23). Under the assumption of a quasi-steady state for $D_A$, $D'_A$, $D_R$, and $D'_R$, we set the derivatives in Eqs. (18)-(21) equal to zero. From this procedure, we deduce that $D'_A = (\theta_A / \gamma_A) D_A$, which, when combined with the conservation law $D'_A + D_A = N_A$, where $N_A$ is the gene copy number, gives the following effective rate equation for $M_A$:

$$dM_A / dt = \left( \alpha'_A \frac{A}{A + c_A} + \alpha_A \frac{c_A}{A + c_A} \right) N_A - \delta_{M_A} M_A. \qquad (28)$$

We see that the effective rate constant has a Hill function form with $c_A = \theta_A / \gamma_A$. Note that the production rate for $M_A$ approaches $\alpha'_A N_A$ for large $A$ ($\gg c_A$) and $\alpha_A N_A$ for small $A$ ($\ll c_A$). The rate equation for $M_R$ becomes

$$dM_R / dt = \left( \alpha'_R \frac{A}{A + c_R} + \alpha_R \frac{c_R}{A + c_R} \right) N_R - \delta_{M_R} M_R, \qquad (29)$$

where $c_R = \theta_R / \gamma_R$. For the parameter values used in this paper, we have $c_A = 50$ and $c_R = 100$. Eq. (28), (29) and Eqs. (24)-(26) yield deterministic results similar to the full VKBL model. For example, the threshold for sustained oscillations is $\delta_R = 0.0878$ in the reduced model and $\delta_R = 0.0877$ in the full model. However, as we shall show, the stochastic properties are somewhat different. As an aside, we note that Vilar et al. (2002) employed a more drastic reduction to just two rate equations for *R* and *C*.

To explore the stochastic properties of the reduced model, we now translate the "reduced" rate equation formulation to an equivalent chemical reaction formulation. Reactions (1)-(4) are assumed to be in quasi-steady state given the current value of *A*. Reactions (5)-(6) and (11)-(12) are replaced by

$$D_A \xrightarrow{\alpha_{Aeff}} D_A + M_A \tag{30}$$

and

$$D_R \xrightarrow{\alpha_{Reff}} D_R + M_A, \tag{31}$$

where

$$\alpha_{Aeff} = \alpha'_A \frac{A}{A+c_A} + \alpha_A \frac{c_A}{A+c_A} \tag{32}$$

and

$$\alpha_{Reff} = \alpha'_R \frac{A}{A+c_R} + \alpha_R \frac{c_R}{A+c_R}. \tag{33}$$

Figure 5 plots the regularity of the noise-induced oscillations as a function of the gene copy number (with $N_A = N_R$), comparing the results for the full stochastic reaction model (repeated from Fig. 2) with those from the reduced model, Eqs. (30)-(31) and (7)-(16), with the same set of parameters. We see that the reduced model fails to capture the stochastic coherence (coherence resonance) effect, that is, the reduced model results do not display a maximum as a function of gene copy number. We conclude that the adiabatic assumption removes some of the important features of the stochastic dynamics even though the usual "folk lore" is that fluctuations in the rapidly changing variables are less important than fluctuations in the slowly changing variables.

As mentioned previously, the VKBL model for the parameter values used here falls in the class of excitable systems. Our analysis indicates that the stochastic dynamics of the system near the deterministic fixed point must include the dynamics on both the fast and slow manifolds. The deterministic behavior is changed only slightly by adiabatic elimination of the fast variables, but the stochastic dynamics near the fixed point can be somewhat different.

## 5. Conclusions

We have demonstrated that stochastic coherence occurs in the dynamics of a gene circuit model in both the stochastic reaction description and in the stochastic rate equation formulation: the regularity of the oscillations exhibits a maximum as a function of noise variance, which for intrinsic noise is linked to system size. The stochastic rate equation model exhibits the distinction between fluctuations in the fast and slow dynamics that had been observed previously in models of excitable neurons (Hilborn and Erwin, 2004; Hilborn and Erwin, 2005). We have also shown that models that eliminate the fast dynamics of the operator binding and unbinding miss some of the important stochastic behavior of the gene circuit. Although the parameters used in the model are generic, they are nevertheless typical. Moreover, the dynamical features of the gene circuit model are also generic, so we expect that similar behavior will occur in other models. Hence, we argue that real gene circuits may in fact take advantage of fluctuations to maintain regular oscillatory behavior. Such ideas might be verified with the use of synthetic gene circuits (Elowitz and Leibler, 2000; Murphy et al., 2007).

**Acknowledgement**

R. C. H. thanks the Center for Theoretical Biological Physics and the Institute for Nonlinear Science at the University of California at San Diego for support for a visit during which this work was begun.

---------

**References**

Barkai, N., and Leibler, S., 2000. Circadian clocks limited by noise. Nature 403, 267-268.

Becker-Weimann, S., Wolf, J., Herzel, H., and Kramer, A., 2004. Modeling Feedback Loops of the Mammalian Circadian Oscillator. Biophys. J. 87, 3023-3034.

Busino, L., Bassermann, F., Maiolica, A., Lee, C., Nolan, P.M., Godinho, S.I.H., Draetta, G.F., and Pagano, M., 2007. SCF$^{Fbx13}$ Controls the Oscillation of the Circadian Clock by Directing the Degradation of Cryptochrome Proteins. Science 316, 900-904.

Dayan, P., and Abbott, L.F., 2001. Theoretical Neuroscience: Computational and Mathematical Modeling of Neural Systems. MIT Press, Cambridge.

Elowitz, M., Levine, A., Siggia, E., and Swain, P., 2002. Stochastic gene expression in a single cell. Science 297, 1183-1186.

Elowitz, M.B., and Leibler, S., 2000. A synthetic oscillatory network of transcriptional regulators. Nature 403, 335-338.

Gillespie, D.T., 1976. A General Method for Numerically Simulating the Stochastic Time Evolution of Coupled Chemical Reactions. J. Comput. Phys. 22, 403-434.

Gillespie, D.T., 1977. Exact Stochastic Simulation of Coupled Chemical Reactions. J. Phys. Chem. 81, 2340-2361.

Gillespie, D.T., 2000. The chemical Langevin equation. J. Chem. Phys. 113, 297-306.

Guido, N.J., Wang, X., Adalsteinsson, D., McMillen, D., Hasty, J., Cantor, C.R., Elston, T., and Collins, J.J., 2006. A bottom-up approach to gene regulation. Nature 439, 856-860.

Hilborn, R.C., and Erwin, R.J., 2004. Coherence resonance in models of excitable neurons with noise in both the fast and slow dynamics. Phys. Lett. A 322, 19-24.

Hilborn, R.C., and Erwin, R.J., 2005. Fokker-Planck analysis of stochastic coherence in models of an excitable neuron with noise in both fast and slow dynamics. Phys. Rev. E 72, 031112-1-14.

Hou, Z.-H., and Xin, H.-W., 2003. Internal noise stochastic resonance in a circadian clock system. J. Chem. Phys. 119, 11508-11512.

Jung, P., and Shuai, J.-W., 2001. Optimal sizes of ion channel clusters. Europhys. Lett. 56, 29-35.

Kærn, M., Blake, W.J., and Collins, J.J., 2003. The Engineering of Gene Regulatory Networks. Annu. Rev. Biomed. Eng. 5, 179-206.

Kærn, M., Elston, T.C., Blake, W.J., and Collins, J.J., 2005. Stochasticity in Gene Expression: From Theories to Phenotypes. Nat. Rev. Genet. 6, 451-464.

Kepler, T.B., and Elston, T.C., 2001. Stochasticity in Transcriptional Regulation: Origins, Consequences, and Mathematical Representations. Biophys. J. 81, 3116-3136.

Lindner, B., García-Ojalvo, J., Neiman, A., and Schimansky-Geier, L., 2004. Effects of noise in excitable systems. Phys. Rep. 392, 321-424.

McAdams, H.H., and Arkin, A., 1997. Stochastic mechanisms in gene expression. Proc. Nat. Acad. Sci. 94, 814-819.

Mettetal, J., Muzzey, D., Pedraza, J.M., Ozbudak, E.M., and van Oudenaarden, A., 2006. Predicting stochastic gene expression dynamics in single cells. Proc. Nat. Acad. Sci. 103, 7304-7309.

Murphy, K.F., Balázsi, G., and Collins, J.J., 2007. Combinatorial promoter design for engineering noisy gene expression. Proc. Nat. Acad. Sci. 104, 12726-31.

Pedraza, J.M., and van Oudenaarden, A., 2005. Noise Progagation in Gene Networks. Science 307, 1965-1969.

Pikovsky, A.S., and Kurths, J., 1997. Coherence Resonance in a Noise-Driven Excitable System. Phys. Rev. Lett. 78, 775-778.

Raser, J.M., and O'Shea, E.K., 2005. Noise in Gene Expression: Origins, Consequences, and Control. Science 309, 2010-2013.

Schmid, G., Goychuk, I., and Hänggi, P., 2001. Stochastic resonance as a collective property of ion channel assemblies. Europhys. Lett. 56, 22-28.

Shuai, J.-W., and Jung, P., 2002a. Optimal Intracellular Calcium Signalling. Phys. Rev. Lett. 88, 068102-1-4.

Shuai, J.-W., and Jung, P., 2002b. Stochastic Properties of $Ca^{2+}$ Release of Inositol 1,4,5-Triphosphate Receptor Clusters. Biophys. J. 83, 87-97.

Shuai, J.-W., and Jung, P., 2005. Entropically Enhanced Excitability in Small Systems. Phys. Rev. Lett. 95, 114501-1-4.

Steuer, R., 2004. Effects of stochasticity in models of the cell cycle: from quantized cycle times to noise-induced oscillations. J. Theor. Bio. 228, 293-301.

van Kampen, N.G., 1992. Stochastic Processes in Physics and Chemistry. Elsevier, Amsterdam.

Vilar, J.M.G., Kueh, H.Y., Barkai, N., and Leibler, S., 2002. Mechanisms of noise-resistance in genetic oscillators. Proc. Nat. Acad. Sci. 99, 5988-5992.

Volfson, D., Marchiniak, J., Blake, W.J., Ostroff, N., Tsimring, L.S., and Hasty, J., 2006. Origins of extrinsic variability in eukaryotic gene expression. Nature 439, 861-864.

Wang, Z.-W., Hou, Z.-H., and Xin, H.-W., 2007. Optimal Internal Noise for Mammalian Circadian Oscillator. Chin. J. Chem. Phys. 20, 119-125.

Wang, Z., Hou, Z., and Xin, H., 2005. Internal noise stochastic resonance of synthetic gene network. Chem. Phys. Lett. 401, 307-311.

Zaikin, A., García-Ojalvo, J., Bascones, R., Ullner, E., and Kurths, J., 2003. Doubly Stochastic Coherence via Noise-Induced Symmetry in Bistable Neural Models. Phys. Rev. Lett. 90, 030601-4.

Zeng, S., and Jung, P., 2004. Mechanism for neuronal spike generation by small and large ion channel clusters. Phys. Rev. E 70, 011903-1-8.

Zhang, J., Hou, Z., and Xin, H., 2004. System-Size Biresonance for Intracellular Calcium Signalling. ChemPhysChem 5, 1041-1045.

**Figure Captions**

Fig. 1. Oscillations in the number $R(t)$ of Repressor proteins in the stochastic reaction model with $\delta_R = 0.06$ and $N_A = N_R = 1$ (that is, one gene copy for each of the two proteins).

Fig. 2. The regularity R of the oscillations in the stochastic reaction model with $\delta_R = 0.06$ plotted as a function of the logarithm of the gene copy number $N_A$, with $N_A = N_R$. Each data point represents the results of averaging five noise realizations, each corresponding to about 4000 hours of simulated dynamics. The uncertainty bars indicate the standard deviation of the mean regularity from the five noise realizations. Squares: parameters as described in the text. Circles: all transcription rates multiplied by 1.5. Triangles: all transcription rates multiplied by 0.5.

Fig 3. Noise-induced oscillations in the number of repressor proteins $R(t)$ from Eqs. (18)–(26) with noise added to Eq. (24) with $\delta_R = 0.06$ and $\log_{10} d_A = 2$ and $D_R(0) = 1$, $D_A(0) = 1$. All other initial species numbers are zero.

Fig. 4. Stochastic coherence in the stochastic differential equation model with $\delta_R = 0.06$. Regularity R plotted as a function of the logarithm of the noise parameter $d$. Each data point is the result of averaging five noise realizations, each simulating 2000 hours of the gene circuit dynamics. Circles: noise added only to Eq. (24). Squares: noise added only to Eq. (25). The uncertainty bars indicate the standard deviation of the mean regularity from the five noise realizations.

Fig. 5. The regularity R plotted as a function of the logarithm of gene copy number with $N_A = N_R$. Squares: the full stochastic reaction model (repeated from Fig. 2) with parameters indicated in the text. Circles: the adiabatic version of the stochastic reaction model. Note that the adiabatic version does not exhibit stochastic coherence.

Figure 1

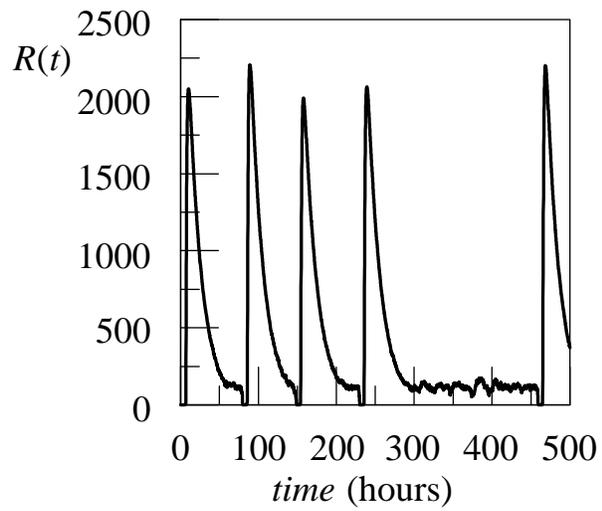

Fig. 1. Oscillations in the number $R(t)$ of Repressor proteins in the stochastic reaction model with $\delta_R = 0.06$ and $N_A = N_R = 1$ (that is, one gene copy for each of the two proteins).

Figure 2

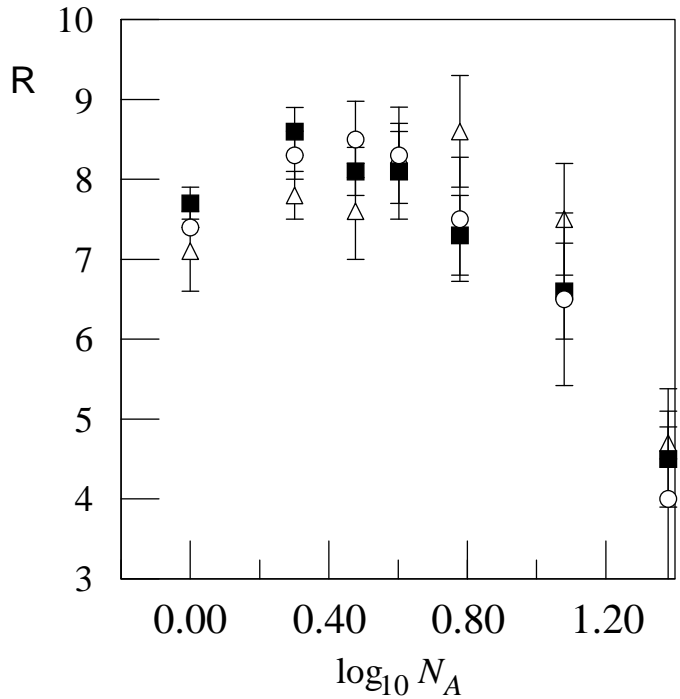

Fig. 2. The regularity R of the oscillations in the stochastic reaction model with $\delta_R = 0.06$ plotted as a function of the logarithm of the gene copy number $N_A$, with $N_A = N_R$. Each data point represents the results of averaging five noise realizations, each corresponding to about 4000 hours of simulated dynamics. The uncertainty bars indicate the standard deviation of the mean regularity from the five noise realizations. Squares: parameters as described in the text. Circles: all transcription rates multiplied by 1.5. Triangles: all transcription rates multiplied by 0.5.

Figure 3

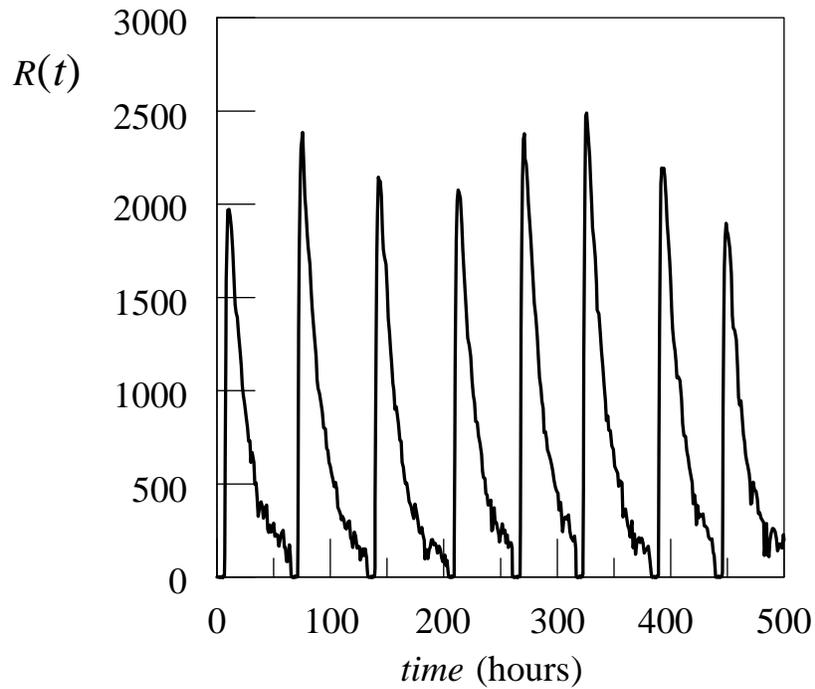

Fig. 3. Noise-induced oscillations in the number of repressor proteins $R(t)$ from Eqs. (18)−(26) with noise added to Eq. (24) with $\delta_R = 0.06$ and $\log_{10} d_A = 2$. Initial conditions: $D_R(0) = 1$, $D_A(0) = 1$. All other initial species numbers are zero.

Figure 4

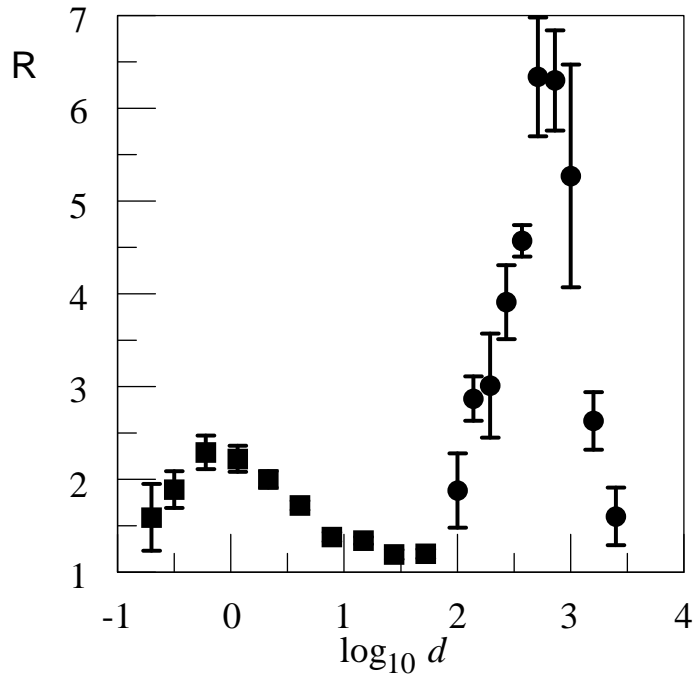

Fig. 4. Stochastic coherence in the stochastic differential equation model with $\delta_R = 0.06$. Regularity R plotted as a function of the logarithm of the noise parameter $d$. Each data point is the result of averaging five noise realizations, each simulating 2000 hours of the gene circuit dynamics. Circles: noise added only to Eq. (24). Squares: noise added only to Eq. (25). The uncertainty bars indicate the standard deviation of the mean regularity from the five noise realizations.

Figure 5

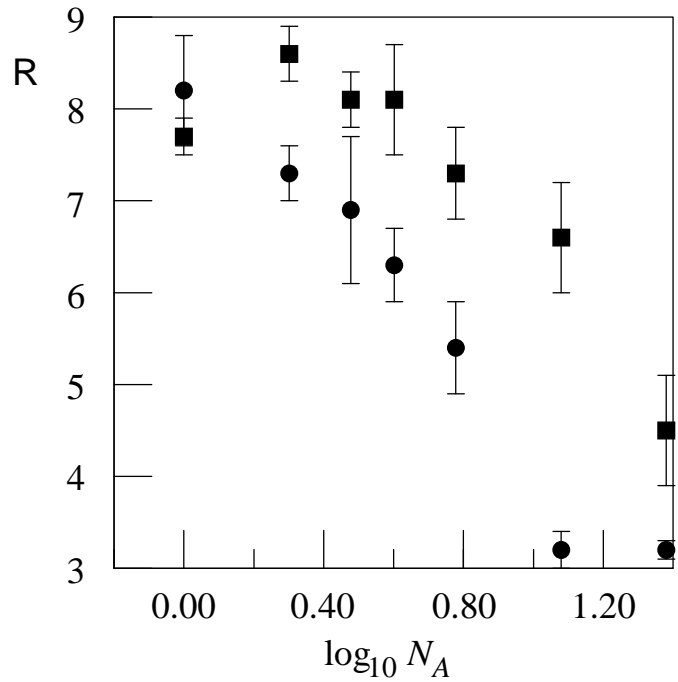

Fig. 5. The regularity R plotted as a function of the logarithm of gene copy number with $N_A = N_R$. Squares: the full stochastic reaction model (repeated from Fig. 2) with parameters indicated in the text. Circles: the adiabatic version of the stochastic reaction model. Note that the adiabatic version does not exhibit stochastic coherence.